\RequirePackage{lineno}
\documentclass[aps,prd,twocolumn,superscriptaddress,amsmath,amssymb,amsart,showpacs]{revtex4}

\usepackage{comment}
\usepackage{multirow}
\usepackage{dcolumn}
\usepackage{bm}
\usepackage{setspace}
\usepackage{graphicx}
\usepackage{amssymb}
\includecomment{printrobustnotes}
\includecomment{printallnotes}
\usepackage{xspace}

\newcommand{\note}[1]{\xspace}
\newcommand{\robustnote}[1]{\xspace}
\begin{printrobustnotes}
\renewcommand{\robustnote}[1]{\textbf{\textit{#1}}\xspace}
\end{printrobustnotes}
\begin{printallnotes}
\renewcommand{\note}[1]{\textbf{\textit{#1}}\xspace}
\renewcommand{\robustnote}[1]{\textbf{\textit{#1}}\xspace}
\end{printallnotes}
\newcommand{\et}     {\ensuremath{E_{T}}\xspace}

\newcommand{\met}    {\ensuremath{\text{\raisebox{.3ex}{$\not$}}\et}\xspace}

\begin{document}


\title{{\bf A Direct Top-Quark Width Measurement at CDF}}

\affiliation{Institute of Physics, Academia Sinica, Taipei, Taiwan 11529, Republic of China} 
\affiliation{Argonne National Laboratory, Argonne, Illinois 60439, USA} 
\affiliation{University of Athens, 157 71 Athens, Greece} 
\affiliation{Institut de Fisica d'Altes Energies, Universitat Autonoma de Barcelona, E-08193, Bellaterra (Barcelona), Spain} 
\affiliation{Baylor University, Waco, Texas 76798, USA} 
\affiliation{Istituto Nazionale di Fisica Nucleare Bologna, $^{bb}$University of Bologna, I-40127 Bologna, Italy} 
\affiliation{Brandeis University, Waltham, Massachusetts 02254, USA} 
\affiliation{University of California, Davis, Davis, California 95616, USA} 
\affiliation{University of California, Los Angeles, Los Angeles, California 90024, USA} 
\affiliation{Instituto de Fisica de Cantabria, CSIC-University of Cantabria, 39005 Santander, Spain} 
\affiliation{Carnegie Mellon University, Pittsburgh, Pennsylvania 15213, USA} 
\affiliation{Enrico Fermi Institute, University of Chicago, Chicago, Illinois 60637, USA}
\affiliation{Comenius University, 842 48 Bratislava, Slovakia; Institute of Experimental Physics, 040 01 Kosice, Slovakia} 
\affiliation{Joint Institute for Nuclear Research, RU-141980 Dubna, Russia} 
\affiliation{Duke University, Durham, North Carolina 27708, USA} 
\affiliation{Fermi National Accelerator Laboratory, Batavia, Illinois 60510, USA} 
\affiliation{University of Florida, Gainesville, Florida 32611, USA} 
\affiliation{Laboratori Nazionali di Frascati, Istituto Nazionale di Fisica Nucleare, I-00044 Frascati, Italy} 
\affiliation{University of Geneva, CH-1211 Geneva 4, Switzerland} 
\affiliation{Glasgow University, Glasgow G12 8QQ, United Kingdom} 
\affiliation{Harvard University, Cambridge, Massachusetts 02138, USA} 
\affiliation{Division of High Energy Physics, Department of Physics, University of Helsinki and Helsinki Institute of Physics, FIN-00014, Helsinki, Finland} 
\affiliation{University of Illinois, Urbana, Illinois 61801, USA} 
\affiliation{The Johns Hopkins University, Baltimore, Maryland 21218, USA} 
\affiliation{Institut f\"{u}r Experimentelle Kernphysik, Karlsruhe Institute of Technology, D-76131 Karlsruhe, Germany} 
\affiliation{Center for High Energy Physics: Kyungpook National University, Daegu 702-701, Korea; Seoul National University, Seoul 151-742, Korea; Sungkyunkwan University, Suwon 440-746, Korea; Korea Institute of Science and Technology Information, Daejeon 305-806, Korea; Chonnam National University, Gwangju 500-757, Korea; Chonbuk National University, Jeonju 561-756, Korea} 
\affiliation{Ernest Orlando Lawrence Berkeley National Laboratory, Berkeley, California 94720, USA} 
\affiliation{University of Liverpool, Liverpool L69 7ZE, United Kingdom} 
\affiliation{University College London, London WC1E 6BT, United Kingdom} 
\affiliation{Centro de Investigaciones Energeticas Medioambientales y Tecnologicas, E-28040 Madrid, Spain} 
\affiliation{Massachusetts Institute of Technology, Cambridge, Massachusetts 02139, USA} 
\affiliation{Institute of Particle Physics: McGill University, Montr\'{e}al, Qu\'{e}bec, Canada H3A~2T8; Simon Fraser University, Burnaby, British Columbia, Canada V5A~1S6; University of Toronto, Toronto, Ontario, Canada M5S~1A7; and TRIUMF, Vancouver, British Columbia, Canada V6T~2A3} 
\affiliation{University of Michigan, Ann Arbor, Michigan 48109, USA} 
\affiliation{Michigan State University, East Lansing, Michigan 48824, USA}
\affiliation{Institution for Theoretical and Experimental Physics, ITEP, Moscow 117259, Russia}
\affiliation{University of New Mexico, Albuquerque, New Mexico 87131, USA} 
\affiliation{Northwestern University, Evanston, Illinois 60208, USA} 
\affiliation{The Ohio State University, Columbus, Ohio 43210, USA} 
\affiliation{Okayama University, Okayama 700-8530, Japan} 
\affiliation{Osaka City University, Osaka 588, Japan} 
\affiliation{University of Oxford, Oxford OX1 3RH, United Kingdom} 
\affiliation{Istituto Nazionale di Fisica Nucleare, Sezione di Padova-Trento, $^{cc}$University of Padova, I-35131 Padova, Italy} 
\affiliation{LPNHE, Universite Pierre et Marie Curie/IN2P3-CNRS, UMR7585, Paris, F-75252 France} 
\affiliation{University of Pennsylvania, Philadelphia, Pennsylvania 19104, USA}
\affiliation{Istituto Nazionale di Fisica Nucleare Pisa, $^{dd}$University of Pisa, $^{ee}$University of Siena and $^{ff}$Scuola Normale Superiore, I-56127 Pisa, Italy} 
\affiliation{University of Pittsburgh, Pittsburgh, Pennsylvania 15260, USA} 
\affiliation{Purdue University, West Lafayette, Indiana 47907, USA} 
\affiliation{University of Rochester, Rochester, New York 14627, USA} 
\affiliation{The Rockefeller University, New York, New York 10065, USA} 
\affiliation{Istituto Nazionale di Fisica Nucleare, Sezione di Roma 1, $^{gg}$Sapienza Universit\`{a} di Roma, I-00185 Roma, Italy} 

\affiliation{Rutgers University, Piscataway, New Jersey 08855, USA} 
\affiliation{Texas A\&M University, College Station, Texas 77843, USA} 
\affiliation{Istituto Nazionale di Fisica Nucleare Trieste/Udine, I-34100 Trieste, $^{hh}$University of Trieste/Udine, I-33100 Udine, Italy} 
\affiliation{University of Tsukuba, Tsukuba, Ibaraki 305, Japan} 
\affiliation{Tufts University, Medford, Massachusetts 02155, USA} 
\affiliation{Waseda University, Tokyo 169, Japan} 
\affiliation{Wayne State University, Detroit, Michigan 48201, USA} 
\affiliation{University of Wisconsin, Madison, Wisconsin 53706, USA} 
\affiliation{Yale University, New Haven, Connecticut 06520, USA} 
\author{T.~Aaltonen}
\affiliation{Division of High Energy Physics, Department of Physics, University of Helsinki and Helsinki Institute of Physics, FIN-00014, Helsinki, Finland}
\author{B.~\'{A}lvarez~Gonz\'{a}lez$^v$}
\affiliation{Instituto de Fisica de Cantabria, CSIC-University of Cantabria, 39005 Santander, Spain}
\author{S.~Amerio}
\affiliation{Istituto Nazionale di Fisica Nucleare, Sezione di Padova-Trento, $^{cc}$University of Padova, I-35131 Padova, Italy} 

\author{D.~Amidei}
\affiliation{University of Michigan, Ann Arbor, Michigan 48109, USA}
\author{A.~Anastassov}
\affiliation{Northwestern University, Evanston, Illinois 60208, USA}
\author{A.~Annovi}
\affiliation{Laboratori Nazionali di Frascati, Istituto Nazionale di Fisica Nucleare, I-00044 Frascati, Italy}
\author{J.~Antos}
\affiliation{Comenius University, 842 48 Bratislava, Slovakia; Institute of Experimental Physics, 040 01 Kosice, Slovakia}
\author{G.~Apollinari}
\affiliation{Fermi National Accelerator Laboratory, Batavia, Illinois 60510, USA}
\author{J.A.~Appel}
\affiliation{Fermi National Accelerator Laboratory, Batavia, Illinois 60510, USA}
\author{A.~Apresyan}
\affiliation{Purdue University, West Lafayette, Indiana 47907, USA}
% ...
\collaboration{CDF Collaboration\footnote{With visitors from $^a$University of Massachusetts Amherst, Amherst, Massachusetts 01003,
$^b$Istituto Nazionale di Fisica Nucleare, Sezione di Cagliari, 09042 Monserrato (Cagliari), Italy,
$^c$University of California Irvine, Irvine, CA  92697, 
$^d$University of California Santa Barbara, Santa Barbara, CA 93106
$^e$University of California Santa Cruz, Santa Cruz, CA  95064,
$^f$CERN,CH-1211 Geneva, Switzerland,
$^g$Cornell University, Ithaca, NY  14853, 
$^h$University of Cyprus, Nicosia CY-1678, Cyprus, 
$^i$University College Dublin, Dublin 4, Ireland,
$^j$University of Fukui, Fukui City, Fukui Prefecture, Japan 910-0017,
$^k$Universidad Iberoamericana, Mexico D.F., Mexico,
$^l$Iowa State University, Ames, IA  50011,
$^m$University of Iowa, Iowa City, IA  52242,
$^n$Kinki University, Higashi-Osaka City, Japan 577-8502,
$^o$Kansas State University, Manhattan, KS 66506,
$^p$University of Manchester, Manchester M13 9PL, England,
$^q$Queen Mary, University of London, London, E1 4NS, England,
$^r$Muons, Inc., Batavia, IL 60510,
$^s$Nagasaki Institute of Applied Science, Nagasaki, Japan, 
$^t$National Research Nuclear University, Moscow, Russia,
$^u$University of Notre Dame, Notre Dame, IN 46556,
$^v$Universidad de Oviedo, E-33007 Oviedo, Spain, 
$^w$Texas Tech University, Lubbock, TX  79609, 
$^x$IFIC(CSIC-Universitat de Valencia), 56071 Valencia, Spain,
$^y$Universidad Tecnica Federico Santa Maria, 110v Valparaiso, Chile,
$^z$University of Virginia, Charlottesville, VA  22906,
$^{aa}$Yarmouk University, Irbid 211-63, Jordan,
$^{ii}$On leave from J.~Stefan Institute, Ljubljana, Slovenia, 
}}
\noaffiliation


\begin{abstract}
 
We present a measurement of the top-quark width in the lepton + jets decay channel of $t\bar{t}$ events produced in $p\bar{p}$ collisions at Fermilab's Tevatron collider and collected by the CDF II detector. From a data sample corresponding to 4.3~fb$^{-1}$ of integrated luminosity, we identify 756 candidate events. The top-quark mass and the mass of the hadronically decaying $W$ boson that comes from the top-quark decay are reconstructed for each event and compared with templates of different top-quark widths~($\Gamma_t$) and deviations from nominal jet energy scale~($\Delta_{JES}$) to perform a simultaneous fit for both parameters, where $\Delta_{JES}$ is used for the {\it in situ} calibration of the jet energy scale. By applying a Feldman-Cousins approach, we establish an upper limit at 95$\%$ confidence level~(CL) of $\Gamma_t <$ 7.6 GeV and a two-sided 68$\%$ CL interval of 0.3 GeV $< \Gamma_t <$ 4.4 GeV for a top-quark mass of 172.5 GeV$/c^2$, which are consistent with the standard model prediction.
\end{abstract}
\maketitle
The top quark is the heaviest known elementary particle, whose large mass results in the largest decay width and hence the shortest lifetime of the quarks in the standard model~(SM). A precise measurement of the top-quark width $\Gamma_t$ is a good test of the standard model, whose prediction at the Born level~\cite{theoryUpdate} is affected by the Quantum Chromodynamics~(QCD) radiative corrections of order 10$\%$~\cite{leadingOrder}, as well as by electroweak corrections~\cite{EW1,EW2}, which are of order 1.5$\%$. The dominant decay mode of the top quark in the SM produces a $W$ boson and a bottom quark~($b$). At leading order the total top-quark width is given by $\Gamma^0_t = |V_{tb}|^2G_Fm_t^3/(8\pi \sqrt{2})$, where $V_{tb}$, $G_F$ and $m_t$ are the Cabibbo-Kobayashi-Maskawa matrix element, Fermi coupling constant and the top-quark mass, respectively.  If we take $|V_{tb}|$ to be unity, the next-to-leading order calculation~\cite{theoryUpdate,leadingOrder} with QCD and electroweak corrections predicts $\Gamma_t$ of 1.3 GeV at a top-quark mass of 172.5 GeV/$c^2$~\cite{tmass_ref} and this corresponds to a lifetime of $5 \times 10^{-25}$ s.

A deviation from the SM could indicate a significant contribution of non-SM particles. Novel top-quark decay modes motivated by the large top-quark mass include decay to a charged Higgs $t\rightarrow b + H^+$~\cite{higgs_1,higgs_2,higgs_3,higgs_4}, decay to its supersymmetric scalar partner stop plus neutralinos~\cite{stop_1,stop_2}, and flavor changing neutral current~(FCNC) top-quark decays~\cite{fcnc}. Therefore, the direct measurement of $\Gamma_t$ is a general way to constrain such processes.  The first direct measurement of $\Gamma_t$ was carried out with an integrated luminosity of 1~fb$^{-1}$ of CDF data in the lepton + jets channel~\cite{widthPRL} and set an upper limit on $\Gamma_t$~$<$~13.1~GeV at 95$\%$ confidence level~(CL), while the result of a recent analysis from the D0 experiment at the Tevatron quotes an indirect top-quark width measurement of $\Gamma_t = 1.99^{+0.69}_{-0.55}$ GeV~\cite{D0width}. In this report of the second direct measurement of $\Gamma_t$, we increase the CDF data set to 4.3~fb$^{-1}$ in the lepton + jets channel, apply a kernel density estimation~(KDE) technique~\cite{kde_1,kde_2} to make templates, determine the jet energy scale~(JES) calibration {\it in situ}, and use new methods for setting and incorporating systematic effects. We set a two-sided bound on the top-quark width at $68\%$ CL for the first time.

CDF~II~\cite{cdfii} is a general-purpose detector located at one of the two collision points along the ring of the Tevatron accelerator. A silicon microstrip tracker and a cylindrical drift chamber in a 1.4 T magnetic field serve as a charged particle tracking system. Electromagnetic and hadronic calorimeters are used to measure the energies of electrons and jets. Outside the calorimeters lie drift chambers which can detect muons. We employ a cylindrical coordinate system for the detector where $\theta$ and $\phi$ are the polar and azimuthal angles, respectively, with respect to the proton beam, and pseudorapidity $\eta \equiv -\ln \tan(\theta/2)$. Transverse energy and momentum are defined as $E_T = E \sin\theta$ and $p_T = p \sin\theta$, respectively, where $E$ and $p$ are energy and momentum.

Top quarks decay almost exclusively to a $W$ boson and a $b$ quark through the weak interaction in the SM. We identify $t\bar{t}$ events in the lepton + jets channel, where one $W$ boson decays to a charged lepton and neutrino, and the other $W$ boson decays to two quarks. The $t\bar{t}$ candidate events used in this analysis are collected by triggers that identify at least one high-$p_T$ lepton. Offline these events are selected by requiring a high-$E_T$ electron or high-$p_T$ muon~($E_T$ or $p_T$ $>$ 20 GeV), large missing transverse energy $\met$~($\met$ $>$ 20 GeV) due to the undetected neutrino from the leptonic $W$ decay, and at least four hadronic jets. Jets are reconstructed with the {\sc jetclu}~\cite{jetclu} cone algorithm using a cone radius of $\Delta R=\sqrt{{\Delta\eta}^2+{\Delta\phi}^2}=0.4$. To determine if a jet comes from a $b$ quark, the {\sc secvtx}~\cite{secvtx} algorithm, which makes use of the transverse decay length of a $b$ quark inside a jet ($b$-tag), is applied. At least one jet must be identified as $b$-tagged. We divide the candidate events into those with one $b$-tagged jet and those with two or more $b$-tagged jets in order to improve the usage of statistical information, since these two kinds of events have different signal-to-background ratios. When an event has one $b$-tagged jet~($b$-jet), we require this event to have exactly four jets each with $E_T >$ 20 GeV; when an event contains two or more $b$-jets, three jets are required to have $E_T >$ 20 GeV, the fourth must have $E_T >$ 12 GeV, and the event is allowed to have extra jets. More details about event selection criteria can be found in~\cite{e_select}.

Monte Carlo~(MC) simulated signal samples are created for a fixed top-quark mass of 172.5 GeV$/c^2$ by the {\sc pythia} version 6.216~\cite{pythia} event generator and have different values of $\Gamma_{t}$ between 0.1 GeV and 30 GeV, as well as various values of $\Delta_{JES}$, which is the difference between the JES effects in MC simulation and data and has a range from -3.0~$\sigma_{c}$ to +3.0~$\sigma_{c}$, where $\sigma_{c}$ is the CDF JES fractional uncertainty~\cite{jes_def}. The overall rate of background events with one $W$ boson and additional jets ($W$ + jets), the dominant background process, is determined using data after subtracting off the rate of events coming from QCD multi-jet production (non-$W$ events), and separating out a MC  based estimate for  electroweak processes~(EWK) such as diboson ($WW/WZ/ZZ$) and single-top production. The fractions of $W$ + jets events with heavy flavor quarks ($Wc$, $Wc\bar{c}$, and $Wb\bar{b}$ events) are determined from MC simulated samples. The rate with which events with a $W$ boson and light flavor quarks contain a misidentified $b$-jet is determined using data samples triggered by the presence of jets. Table~\ref{table_background} summarizes the background compositions, and the selection criteria for determining the background rates are described in~\cite{bkgd_sel_ref}. Diboson backgrounds are modeled with {\sc pythia} version 6.216~\cite{pythia} and $W$ + jets by {\sc alpgen} version 2.10$^{'}$~\cite{alpgen}, with jet fragmentation modeled by {\sc pythia} version 6.325~\cite{pythia}. Single-top production events are generated by {\sc madevent}~\cite{madevent} and their fragmentation is modeled with {\sc pythia} version 6.409~\cite{pythia_2}. 
%
%

We use a template method to extract $\Gamma_t$. Two observables, the reconstructed top-quark mass~($m_t^{reco}$) and the invariant mass of the two jets from the hadronically decaying $W$ boson~($m_{jj}$), are built for each data event or MC simulated event~(both signal and background). With the assumption that the leading~(highest $E_T$) four jets in the detector come from the four primary quarks of $t\bar{t}$ events in lepton + jets channel, there are 12 possible assignments of jets to quarks in each event. The neutrino transverse momentum is calculated from the imbalance of the transverse momentum of decaying products, jets and lepton, with unclustered energy taken into account, which is the energy in the calorimeter not associated with the lepton or one of the four leading jets. We use a $\chi^2$-like kinematic fitter~\cite{tmt_mass} to fit the top-quark mass for each assignment, assuming the mass equality of the top and anti-top quarks, and take $m_t^{reco}$ from the assignment that has the lowest $\chi^2$. Events with $\chi^2 > 9.0$ are removed from the sample to reject poorly reconstructed events. We also apply boundary cuts on $m_{t}^{reco}$~(110 GeV$/c^2$ $< m_{t}^{reco} <$ 350 GeV$/c^2$) and $m_{jj}$~(50 GeV$/c^2$ $< m_{jj} <$ 115 GeV$/c^2$ for single $b$-tag events and 50 GeV$/c^2$ $< m_{jj} <$ 125 GeV$/c^2$ for double $b$-tag events) and normalize the probability density functions~(p.d.f.) in these regions. The di-jet mass $m_{jj}$ is calculated as the invariant mass of two non-$b$-tagged jets which provides the closest value to the world average $W$ boson mass of 80.40 GeV$/c^2$~\cite{w_mass}.  The estimated number of background events and observed number of events from a data set corresponding to an integrated luminosity of 4.3~fb$^{-1}$  after event selection, $\chi^2$ cut, and boundary cuts are listed in Table~\ref{table_background}.
\begin{table}
\begin{center}
\caption[Expected number of Lepton+Jets background events]{The sources
and expected numbers of background events in the lepton + jets channel, and the number of events observed for single $b$-tag and double $b$-tag samples after event selection, $\chi^2$ cut, and boundary cuts.}
\begin{ruledtabular}
\begin{tabular}{ccc}
\label{table_background}
            &single $b$-tag         &double $b$-tag\\
\hline
$W$ + jets & 85.6 $\pm$ 21.8 & 9.8 $\pm$ 2.9\\
non-$W$ & 24.5 $\pm$ 20.6 & 2.4 $\pm$ 1.8\\
EWK & 10.2 $\pm$ 0.8 & 2.4 $\pm$ 0.2\\

\hline
Total background&120.2 $\pm$ 30.0 &14.6 $\pm$ 3.4\\
Observed events & 542 & 214 
\end{tabular}
\end{ruledtabular}
\end{center}
\end{table}
\begin{figure}
\includegraphics[width=0.48\textwidth]{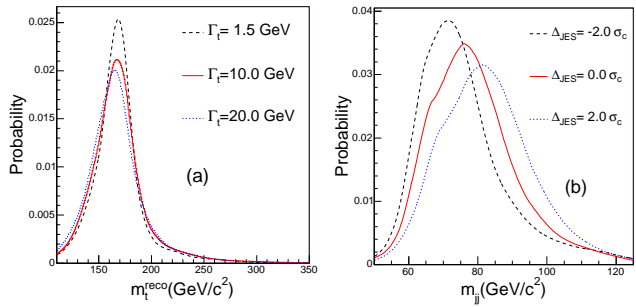}
\caption[tmass templates in MC]{ (a) Probability density functions of $m_t^{reco}$  from double b-tag events for MC simulated samples of different values of $\Gamma_t$; (b)  p.d.f.'s of $m_{jj}$ from double b-tag events for MC simulated samples of different values of $\Delta_{JES}$. }
\label{fig_mass_wjj}
\end{figure}
%
After event reconstruction, we use the MC simulated models of signal and background processes to build two-dimensional p.d.f.'s that give the probabilities of observing a pair of values of $m_t^{reco}$ and $m_{jj}$, given $\Gamma_t$ and $\Delta_{JES}$. We employ a KDE that associates to each data point a function~(called a kernel function) and uses a non-parametric method to estimate the p.d.f.'s of a variable by summing all the kernel functions, without any assumption about the functional form of the p.d.f.'s. Figure~\ref{fig_mass_wjj} shows the p.d.f.'s of $m_t^{reco}$ with different $\Gamma_t$ and the $m_{jj}$ with various $\Delta_{JES}$ from a full MC simulation. We compare the distributions of data with signal and background p.d.f.'s using an unbinned maximum likelihood fit~\cite{likelihood_fit}, where the likelihood function ${\mathcal{L}}$ is the same as what is defined in~\cite{kde_2} except that the fitting parameters are now $\Gamma_t$ and $\Delta_{JES}$. We minimize the negative logarithm of the likelihood using {\sc minuit}~\cite{minuit} to extract the top-quark width. The fitting to $\Delta_{JES}$ reduces the JES systematic effect on $\Gamma_t$ and thus improves the sensitivity to the top width.   

We set the limit(s) on $\Gamma_t$ via the Feldman-Cousins method~\cite{fc} which determines the confidence intervals. The ordering parameter for MC simulated samples that appears in~\cite{fc} is defined here as $\Delta\chi^2 \equiv \chi^2_{input} - \chi^2_{min}$, where $\chi^2 = -2\log({\mathcal{L}})$~(different from the $\chi^2$ mentioned in event reconstruction), $\chi^2_{min}$ is the minimal $\chi^2$ value and $\chi^2_{input}$ is the $\chi^2$ at the real value of parameters $\Gamma_t$ and $\Delta_{JES}$ of the MC simulated sample. We project the  likelihood function $\mathcal{L}$ onto the $\Gamma_t$ axis~\cite{chi2_proj}. For each value of $\Gamma_t$ we run 6000 pseudo-experiments that generate a distribution of $\Delta\chi^2$ from which we calculate a critical value $\Delta\chi^2_c$ so that 95$\%$ of the pseudo-experiments have a $\Delta\chi^2$ falling in the interval $[0, \Delta\chi^2_c]$. With MC simulated samples of 21 different top widths $\Gamma_t$ we get a profile of $\Delta\chi^2_c(\Gamma_t)$. When analyzing the data we obtain $\Delta\chi^2(\Gamma_t | \text{data}) \equiv -2\log({\mathcal{L}}) + 2\log({\mathcal{L}}_0)$, where ${\mathcal{L}}_0$ is the maximum likelihood value of data fitting, then $\Delta\chi^2(\Gamma_t | \text{data})$ is compared with $\Delta\chi^2_c(\Gamma_t)$ and the accepted interval of $\Gamma_t$ is all points such that $\Delta\chi^2(\Gamma_t | \text{data}) < \Delta\chi^2_c(\Gamma_t)$. From the above method we obtain a purely statistical  upper limit on $\Gamma_t$ at 95$\%$ CL, $\Gamma_t < 6.7$ GeV and a two-sided limit of 0.5 GeV $< \Gamma_t <$ 3.9 GeV at 68$\%$ CL. 
\begin{table}
\begin{center}
\caption[Summary of systematics]{Summary of changes in measured $\Gamma_t$ due to systematic effects.}
\begin{ruledtabular}
\begin{tabular}{cc}
\label{systtablesummary}
Systematic Sources& $\Delta\Gamma_{top}$ (GeV) \\
\hline
Jet energy resolution         & 1.1 \\
Color Reconnection     &0.9 \\
Generator             & 0.4 \\
Residual JES           & 0.3 \\
Parton distribution functions                   &0.3 \\
Multiple Hadron Interaction                 &0.3 \\
gluon gluon fraction            &0.3 \\
Initial/Final state radiation              &0.2 \\
Lepton energy scale         &0.2 \\
$b$ jet energy           &0.2 \\
Background shape       &0.1 \\
\hline                      
Total systematic effect       &1.6\\
\end{tabular}
\end{ruledtabular}
\end{center}
\end{table}

We examine systematic effects by comparing MC simulated experiments in which we float parameters within their uncertainties. As seen from Table~\ref{systtablesummary}, the dominant systematic effects come from jet energy resolution and color reconnection (CR)~\cite{cr_2,cr_3}, which is a rearrangement of the underlying color structure of an event from its simplest configuration. For the jet energy resolution effect, we compare jet energy resolution between data and MC simulated samples using one photon + one jet events and smear jet energy with the difference between data and MC simulated samples. We study the effect of CR by using {\sc pythia} version 6.4 with different tunes (with and without CR) and evaluate the difference. The systematic effect due to JES is very small because we perform an {\it in situ} JES calibration. Other smaller systematic effects include those due to the MC generator, the parton distribution functions, and multiple hadron interactions, details of which can be found in ~\cite{tmass_ref,cr_1}. The total change of measured $\Gamma_t$  due to these systematic effects is 1.6 GeV. We studied the dominant systematic uncertainties by varying top-quark width, and found no significant dependence of systematic effects on different top-quark widths.

%
%
\begin{figure}
\includegraphics[width=0.4\textwidth]{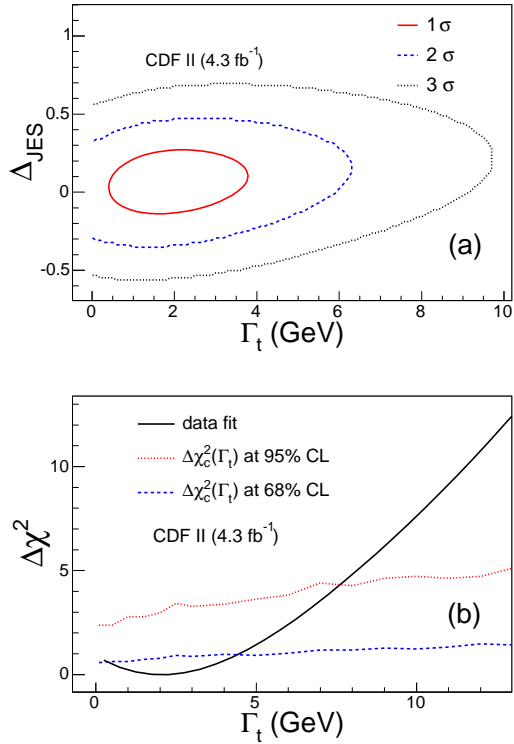}
\caption[data Fit]{ (a) Contours of the two-dimensional negative log likelihood function from data fit. The three different contours represent different values of $-\log({\mathcal{L}})$: 0.5, 2.0, and 4.5. Systematic effects are not included here. (b) Overlap of the $\Delta\chi^2_c(\Gamma_t)$ profile and the data fit that comes from projection of the two-dimensional data fit onto the $\Gamma_t$ axis, the intersection of which gives a limit(s) on $\Gamma_t$. Systematic effects are included in the plots, both for 68$\%$ and 95$\%$ CL.}
\label{fig_fc_datafit}
\end{figure}
To incorporate systematic effects into the limit(s) on $\Gamma_t$ we use a convolution method for folding systematic effects into the likelihood function~\cite{incor_syst,incor_syst_2}. We convolve the likelihood function with a Gaussian p.d.f. that has a width equal to 1.6 GeV and is centered at 0. With this new likelihood function we apply the Feldman-Cousins approach and find an upper limit of $\Gamma_t < 7.6$ GeV at 95$\%$ CL. Using the same approach we are also able to set a two-sided bound for $\Gamma_t$ at 68$\%$ CL: 0.3 GeV $< \Gamma_t <$ 4.4 GeV. Figure~\ref{fig_fc_datafit}(a) shows the data fit from the two-dimensional likelihood function with the statistical uncertainty. The overlap of the $\Delta\chi^2_c(\Gamma_t)$ profile and the one-dimensional data fit that comes from the projection of the two-dimensional likelihood function is shown in Fig.~\ref{fig_fc_datafit}(b), on which the point(s) of interception gives the limit(s) of $\Gamma_t$.

In conclusion, a top-quark width measurement in the lepton + jets channel is presented. Using a data set corresponding to an integrated luminosity of 4.3~fb$^{-1}$ collected by CDF and an {\it in situ} JES calibration, we set an upper limit $\Gamma_t <$ 7.6 GeV at 95$\%$ CL assuming a top-quark mass $M_{top} =$ 172.5 GeV$/c^2$, which is consistent with the standard model. We also quote 0.3 GeV $< \Gamma_t <$ 4.4 GeV at 68$\%$ CL, which corresponds to a life time of 1.5 $\times 10^{-25}$ s $< \tau_t <$ 2.2 $\times 10^{-24}$ s. For a typical quark hadronization time scale of 3.3 $\times 10^{-24}$ s~(corresponding to 200 MeV)~\cite{had_decay1,had_decay2}, our result supports top-quark decay before hadronization.

We thank the Fermilab staff and the technical staffs of the participating institutions for their vital contributions. This work was supported by the U.S. Department of Energy and National Science Foundation; the Italian Istituto Nazionale di Fisica Nucleare; the Ministry of Education, Culture, Sports, Science and Technology of Japan; the Natural Sciences and Engineering Research Council of Canada; the National Science Council of the Republic of China; the Swiss National Science Foundation; the A.P. Sloan Foundation; the Bundesministerium f\"ur Bildung und Forschung, Germany; the World Class University Program, the National Research Foundation of Korea; the Science and Technology Facilities Council and the Royal Society, UK; the Institut National de Physique Nucleaire et Physique des Particules/CNRS; the Russian Foundation for Basic Research; the Ministerio de Ciencia e Innovaci\'{o}n, and Programa Consolider-Ingenio 2010, Spain; the Slovak R\&D Agency; and the Academy of Finland. 


\end{document}